\documentclass[11pt,a4paper]{article}

\usepackage{amsmath,amsfonts,amssymb,amsthm,cite}

\evensidemargin=\oddsidemargin
\advance\topmargin by -\headheight
\advance\topmargin by -\headsep

\newtheorem{thm}{Theorem}[section]
\newtheorem{lem}[thm]{Lemma}
\newtheorem{prop}[thm]{Proposition}

\newcommand{\A}{\mathcal{A}}        
\renewcommand{\a}{\alpha}           
\DeclareMathOperator{\ad}{ad}       
\newcommand{\Aun}{\widetilde{\mathcal{A}}} 
\newcommand{\B}{\mathcal{B}}        
\newcommand{\braket}[2]{\langle#1\mathbin|#2\rangle} 
\newcommand{\C}{\mathbb{C}}         
\newcommand{\cc}{\mathbf{c}}        
\newcommand{\D}{\mathcal{D}}        
\newcommand{\del}{\partial}         
\DeclareMathOperator{\Dom}{Dom}     
\newcommand{\Dslash}{{D\mkern-11.5mu/\,}} 
\newcommand{\eps}{\varepsilon}      
\newcommand{\ga}{\gamma}            
\renewcommand{\H}{\mathcal{H}}      
\newcommand{\half}{\tfrac{1}{2}}    
\newcommand{\Ht}{{\widetilde{\mathcal{H}}}} 
\DeclareMathOperator{\Junk}{Junk}   
\DeclareMathOperator{\Ker}{Ker}
\renewcommand{\L}{\mathcal{L}}
\newcommand{\La}{\Lambda}           
\newcommand{\la}{\lambda}           
\newcommand{\mop}{\mathbin{\star_{_\theta}}} 
\newcommand{\N}{\mathbb{N}}         
\newcommand{\Oh}{\mathcal{O}}       
\newcommand{\opp}{{\mathrm{op}}}    
\newcommand{\ox}{\otimes}           
\newcommand{\oxyox}{\otimes\cdots\otimes} 
\newcommand{\pa}{\partial}          
\newcommand{\pd}[2]{\frac{\partial#1}{\partial#2}} 
\newcommand{\R}{\mathbb{R}}         
\newcommand{\roundbraket}[2]{(#1\mathbin|#2)} 
\newcommand{\sepword}[1]{\quad\mbox{#1}\quad} 
\newcommand{\set}[1]{\{\,#1\,\}}     
\DeclareMathOperator{\spec}{spectrum}     
\renewcommand{\SS}{\mathcal{S}}     
\DeclareMathOperator{\Sup}{Sup}
\DeclareMathOperator{\Span}{Span}
\renewcommand{\th}{\theta}          
\newcommand{\tpi}{{\tilde\pi}}      
\DeclareMathOperator{\Tr}{Tr}       
\DeclareMathOperator{\tr}{tr}       
\newcommand{\8}{\bullet}            
\renewcommand{\.}{\cdot}            

\begin{document}

\thispagestyle{empty}

\begin{center}

CENTRE DE PHYSIQUE TH\'EORIQUE$^1$\\
CNRS--Luminy, Case 907\\
13288 Marseille Cedex 9\\
FRANCE\\

\vspace{3cm}

{\Large\textbf{The action functional for Moyal planes}} \\

\vspace{1cm}

{\large Victor. GAYRAL$^2$} \\

\vspace{2cm}

{\large\textbf{Abstract}}
\end{center}

Modulo some natural generalizations to noncompact spaces, we show in this letter
that Moyal planes are nonunital spectral triples in the sense of Connes.
The action functional of these triples is computed, and we obtain the expected result,
ie the noncommutative Yang-Mills action associated with the Moyal product. In
particular, we show that Moyal gauge theory naturally fit into the rigorous framework
of noncommutative geometry.

\vspace{2cm}

\noindent
PACS numbers: 11.10.Nx, 02.30.Sa, 11.15.Kc\\
MSC--2000 classes: 46H35, 46L52, 58B34, 81S30 \\
CPT--03/P.4558
hep-th/0307220
\vspace{4cm}

\noindent $^1$ Unit\'e Propre de Recherche 7061\\
$^2$ Also at Universit\'e de Provence, gayral@cpt.univ-mrs.fr.

\newpage
\section{Introduction}
In the noncommutative field theory picture, prior to quantization, fields are
elements of some function algebra on some manifold endowed with a
noncommutative product, such as the Moyal product. Playing this game
(exchanging the pointwise product for a noncommutative one) is equivalent to
consider field theory on some objects which are no longer smooth manifolds
(noncommutativity destroys the concept of point). The spectral triple
formalism~\cite{Book,ConnesReal,Polaris} is exactly the mathematical and conseptual
framework in which one can rigorously deal with these noncommutative spaces.

More precisely, a noncommutative Riemannian spin geometry is defined by a
triplet $(\A,\H,$$\D)$, where $\A$ is an algebra faithfully represented on some
Hilbert space $\H$ and $\D$ is an unbounded selfadjoint operator. The
commutative case of a compact Riemannian spin manifold, can be recovered with
$\A=C^\infty(M)$, $\H$ the space of $L^2$-sections of the spinor bundle and
$\D$ the Dirac operator~\cite{Polaris}. We may quote~\cite{Stro} for an
approach to the semi-Riemannian case and~\cite{Selene,RennieLocal} for early
approaches to noncompact (i.e., nonunital) noncommutative geometry.

Moyal geometry, based on spaces of functions or distributions endowed with the
Moyal product, is thoroughly investigated in~\cite{nous}. It has been proved
that one can construct nonunital spectral triples with this product. Here we
focus on the construction of the action
functional~\cite{ConnesAction,ConnesL1}, associated with the (below defined)
Moyal spectral triple. Such action is closely related to the spectral triples
formalism, and allows to extract  physically relevant (noncommutative)
geometrical information. We will find that the action functional of the Moyal
plane turns out to be the noncommutative Yang-Mills action of Moyal gauge
theory, with Lagrangean $L(x)=(F^{\mu\nu}\mop F_{\mu\nu}) (x)$, where
$F^{\mu\nu}=\pa^\mu A^\nu-\pa^\nu A^\mu+[A^\mu,A^\nu]_{\mop}$.

In the first part of this letter, a definition of nonunital spectral triple
will be given and discussed, then we will give basic tools and results on
Moyal analysis to construct the nonunital Moyal triple. The last part will be
devoted to the construction and the computation of the action functional.

\section{Nonunital spectral triples}

We write down axioms describing non necessarily compact non commutative spin
geometries. In~\cite{nous}, it is described how one can arrive to them by
learning from examples. In the operatorial setting, such noncommutative spaces
(nonunital spectral triples) are given by the data $(\A,\Aun,\H,\D,J,\chi)$,
where $\chi=1$ in the odd case. $\A$ is an a priori nonunital noncommutative
algebra, $\Aun$ is one (appropriately chosen) of its unitizations, and both are
faithfully represented by bounded operators on an Hilbert space $\H$ (via a
representation $\pi$ ). $\D$ is an unbounded sefladjoint operator such that each
$[\D,a]$, for
$a\in\Aun$ extends to a bounded operator. The triple is said even if there is on $\H$ a
$\mathbb{Z}_2$-grading operator $\chi$ which commutes with $\Aun$ and anticommutes with
$\D$. $J$ is an antiunitary operator having definite commuting properties with respect
to $\chi$ and $\D$, depending on the dimension. Those data $(\A,\Aun,\H,\D,J,\chi)$
must moreover satisfy the following axioms:

\begin{enumerate}
\addtocounter{enumi}{-1}
\item\textit{Compactness}:

For all $a \in \A$ and $\la \notin \spec(\D)$, the operator
$\pi(a)(\D - \la)^{-1}\in\mathcal{K}(\H)$, where $\mathcal{K}(\H)$
is the set of compact operators on the Hilbert space $\H$.

\item\textit{Spectral dimension}:

The spectral dimension $k$ is the unique positive integer such that
$$\pi(a)(|\D| + \eps)^{-1}\in \L^{(k,\infty)}(\H)$$ and
$$\Tr_\omega (\pi(a)(|\D| + \eps)^{-k}) <\infty,$$ and not identically zero,
for all $a$ belonging to a dense ideal of $\A$; $k$ must have the same parity
as the triple.

Here $\L^{(p,\infty)}(\H)$ is the $p$-th weak-Schatten class and $\Tr_\omega$
is any Dixmier trace. Recall that $\L^{(1,\infty)}(\H)$ is the natural domain
of those traces.

\item\textit{Regularity}:

For all $a \in \Aun$, the bounded operators $\pi(a)$ and $[\D,\pi(a)]$
lie in the smooth domain of the derivation $\delta(.)= [|\D|,.]$.

\item\textit{Finiteness}:

The algebras $\A$ and $\Aun$ are pre-$C^*$-algebras. The space
$\H^\infty := \bigcap_{n\in\N} \Dom(\D^n)$ is the $\A$-pullback~\cite{RennieLocal}
of a finite projective $\Aun$-module. Moreover, an
$\A$-valued hermitian structure $\roundbraket{\.}{\.}$ is implicitly defined on
$\H^\infty$ with the noncommutative integral as follows:
$$
\Tr_\omega\bigl( \roundbraket{\pi(a)\xi}{\eta}(|\D| + \eps)^{-k} \bigr)
= \braket{\eta}{\pi(a)\xi},
$$
where $a\in\Aun$ and  $\braket{\.}{\.}$ denotes the standard inner product on~$\H$.

\item\textit{Reality}:

The antiunitary operator $J$ must define a commuting representation on $\H$:
$$[\pi(a), J\pi(b^*)J^{-1}] = 0$$ for all $a,b\in\Aun$. Moreover, $J^2 = \pm 1$
and $J\D = \pm \D J$, and also $J\chi = \pm \chi J$ in the even case, where the
signs depend only on $k\bmod 8$ (see~\cite{ConnesReal} or~\cite{Polaris} for
the table of signs).

\item\textit{First order}:

For all $a,b \in \Aun$, we also have: $[[\D,\pi(a)], J\pi(b^*)J^{-1}] = 0.$

\item\textit{Orientation}:

There is a \textit{Hochschild $k$-cycle} $\cc$, on~$\Aun$ with values in
$\Aun \ox \Aun^\opp$, given by $\cc =\sum_{i\in I}(a^i_0 \ox b^i_0) \ox
a^i_1\oxyox a^i_k$, where $I$ is a finite set and such that
$$
\sum_{i\in I}\pi(a^i_0) J \pi(b^i_0)^*J^{-1} [\D,\pi(a^i_1)]
\cdots [\D,\pi(a^i_k)]=\chi.
$$
\end{enumerate}

These axioms generalize those of the unital (compact)
case~\cite{ConnesReal,Polaris}. Postulates 2, 4, 5 have not been modified from
the compact case, except than in them $\Aun$ has been substituted for $\A$
everywhere. The modifications of the axioms 0 and 1 are quite natural; they
allows us to recover compact operators (recall that in the noncompact manifold
case, pseudodifferential operators of strictly negative order are no longer
compact). We refer to~\cite{nous} on the question why in Axioms~3 and~6 we
need an unitization of the algebra (a compactification of the underlying
"noncommutative space").

\section{The Moyal spectral triple}
\subsection{Basic Moyal analysis}
\par
In this part, a few basic facts on Moyal analysis are given; a complete
review can be found in~\cite{nous}. The starting point is the Moyal product in
its integral form. In the general framework, Moyal products can be defined on
$\R^k$ for a given $k\times k$ skew symmetric matrix $\Theta$:
\begin{eqnarray}
\label{mogeneral}
(f \times_\Theta g)(x):=(2\pi)^{-k}\int  e^{i\xi(x-y)}
f(x-\textstyle{\frac{1}{2}}\Theta \xi)g(y)\;d^ky\; d^k\xi.
\end{eqnarray}
For technical reasons, only the nondegenerate ``Darboux" case will be
considerated: $k=2N$, $\Theta=\th S:=\th \begin{pmatrix} 0 & 1_N \\ -1_N & 0
\end{pmatrix}$. In this case the Moyal product can be rewritten in a more
symmetric way:
\begin{equation}
(f \mop g)(x) := (\pi\th)^{-2N}\int f(y) g(z)\,
e^{\frac{2i}{\th}(x-y)\,\.\,S(x-z)}\;  d^{2N}y \;d^{2N}z .
\label{eq:moyal-prod}
\end{equation}
For $f,g$ lying in the Schwartz space (with its usual topology) this product
is well defined, and with the complex conjugation as involution
$(\SS(\R^{2N}),\mop)$ closes to an $*$-algebra:
\begin{thm}{\rm\cite{Phobos, Deimos}}
$\B_\th:=(\SS(\R^{2N}),\mop)$ is a nonunital, associative, involutive
Fr\'echet algebra with a jointly continuous product.
\end{thm}

This algebra has some (well known) nice algebraic properties, which will be
of practical use to compute the action functional. In particular,
the Leibniz rule is satisfied:
\begin{eqnarray}
\label{leibniz}
\pa^\mu(f\mop g)=\pa^\mu f\mop g+f\mop\pa^\mu g,
\end{eqnarray}
the pointwise product by coordinate functions obeys:
\begin{eqnarray}
\label{poduit}
x^\mu (f \mop g)
= f \mop (x^\mu g) +{\textstyle \frac{i\,\th}{2}} \pd{f}{(Sx)^\mu} \mop g
= (x^\mu f) \mop g - {\textstyle \frac{i\,\th}{2}} f \mop \pd{g}{(Sx)^\mu},
\end{eqnarray}
the ordinary integral is a faithful trace:
\begin{eqnarray}
\label{trace}
\int\; (f\mop g)(x)\; d^{2N}x=\int \; f(x) g(x)\;d^{2N}x=\int \;
(g\mop f)(x)\;d^{2N}x.
\end{eqnarray}
$(\SS(\R^{2N}),\mop)$ has an interesting countable dense subalgebra
consisting of the finite linear combinations of the Wigner eigentransitions.
These particular elements $\{f_{mn}\}_{m,n\in\N^N}$ of $\SS(\R^{2N})$ are
constructed from the N-dimensional harmonic oscillator theory. For
$m,n\in\N^N$, we use the usual multiindex notation: $|m|=\sum_{j=1}^N
m_j$,
$m!=\prod_{j=1}^N m_j!$,
$\delta_{mn}=\prod_{j=1}^N\delta_{m_j n_j}$. The $\{f_{mn}\}$ are defined by:
\begin{equation}
f_{mn}
:= \frac{1}{\sqrt{\th^{|m|+|n|}\,m!n!}}\,(\bar{a})^m \mop f_0 \mop a^n,
\label{eq:basis}
\end{equation}
where the creation and annihilation functions are respectively
\begin{equation}
\bar{a}_l := {\textstyle\frac{1}{\sqrt{2}}} (x_l - ix_{l+N})  \sepword{and}
a_l := {\textstyle\frac{1}{\sqrt{2}}} (x_l + ix_{l+N}),
\label{eq:crea-annl}
\end{equation}
$f_0$ is the Gaussian function $f_0(x) := 2^N e^{-2H/\th}$
associated with the N-dimensional oscillator Hamiltonian function
$H(x):=\frac{1}{2}\sum_{j=1}^N(x_j^2+x_{j+N}^2)$. The main
properties of these functions are summarized in the following
Lemma.

\begin{lem} {\rm\cite{Phobos, Deimos}}
\label{fmn}
Let $m,n,k,l\in\N^N$, then the Wigner eigentransitions satisfy:
$f_{mn}\mop f_{kl}=\delta_{nk}f_{ml}$, $f^*_{mn}=f_{nm}$,
$\langle f_{mn}, f_{kl}\rangle=(2\pi\th)^N\delta_{mk}\delta_{nl}$.
Here $\langle .,.\rangle$ is the standard inner product of
$L^2(\R^{2N})$, in particular
$\{f_{mn}\}\subset\SS(\R^{2N})\subset L^2(\R^{2N})$
is an orthogonal basis.
\end{lem}

With these properties, the $\{f_{mn}\}$ basis allows us to define
the Moyal product on larger spaces than $\SS(\R^{2N})$. For instance,
$f\mop g\in L^2(\R^{2N})$ if $f,g\in L^2(\R^{2N})$ and $(L^2(\R^{2N}),
\mop)$ is an involutive Banach algebra. This can be shown expanding
$f$ and $g$ with respect to the $\{f_{mn}\}$ basis and applying the
Cauchy-Schwarz inequality. As a consequence, if we denote by
$L^\th_f$ the operator of left Moyal multiplication by $f$, acting on
$L^2(\R^{2N})$, we have:
\begin{lem}
\label{rep}
Let $f\in L^2(\R^{2N})$ then
$\|L^\th_f\|\leq(2\pi\th)^{-N/2}\|f\|_2$.
\end{lem}

\subsection{The triple}

Next, a triple is given satisfying all the axioms spelled out in Section~2.
Such choice can't be absolutely unique; however the analytical content of the
data $(\A,\Aun,\H,\D,J,\chi)$, is severely constrained by the axioms. The
choice of the algebra is almost exclusively guided by the finiteness
condition; in other words the Dirac operator tells us what the ``good"
algebra must be. For its unitization, there is an upper bound given by the
boundness of its representation; and the orientation condition gives a lower
bound because we have to be able to construct a good Hochschild cycle with it.
The constraints on the algebras given by the regularity condition are
relatively weak.

The Moyal spectral triple consists of the algebra $\A_\th:=(\D_{L^2},\mop)$,
where $\D_{L^2}$ (see~\cite{Schwartz}) is the space of smooth functions in
$L^2(\R^{2N})$ together with all of their derivatives. Using the Leibniz
rule~(\ref{leibniz}) and estimates similar as in Lemma~\ref{rep}, one can show
that $(\D_{L^2},\mop)$ is an algebra with continuous product. Its unitization
is chosen to be $\Aun_\th:=(\Oh_0,\mop)$, where $\Oh_0$ is the space of
smooth functions with all their derivatives bounded. There is a natural
Fr\'echet topology on $\Oh_0$ given by the seminorms $q_m(f)=\Sup\{|\pa^\a f|,
|\a|\leq m\}$ ---this topology actually comes from the regularity axiom
(see~\cite{RennieLocal}), but is not the only one to be
considered~\cite{Schwartz}. It was shown some time ago~\cite{Amalthea} that
$(\Oh_0,\mop)$ is a unital algebra with a jointly continuous product. For both
$\A_\th$ and $\Aun_\th$, $\B_\th:=(\SS(\R^{2N}),\mop)$ is an essential two
sided ideal.

Those algebras are faithfully represented on $\H:=L^2(\R^{2N})\ox \C^{2^N}$ by
$\pi^\th(f):=L^\th_f\ox 1_{2^N}$. By Lemma~\ref{rep}, $\pi^\th(f)$ is bounded
for $f\in\A_\th$ and for $f\in\Aun_\th$, $L^\th_f$ is by
formula~(\ref{mogeneral}) a zero order pseudodifferential operator ($\Psi$DO)
and hence bounded. The unbounded operator $\D$ is chosen to be the usual Dirac
$\Dslash$ operator of the 2N-dimensional Euclidean flat space,
$\Dslash=-i\pa_\mu\ox\ga^\mu$. Here the $\ga^\mu$ are the Clifford matrices
associated to $(\R^{2N},\eta)$ in their irreducible
$2^N$-dimensional representation and $\eta$ is the standard Euclidean metric
on $\R^{2N}$. For $f\in\A_\th$ or $f\in\Aun_\th$, the Leibniz
rule~(\ref{leibniz}) immediately shows that, $[\Dslash,\pi^\th(f)]
=-iL^\th_{\pa_\mu f}\ox\ga^\mu$, which is bounded, too. As the triple is even,
the grading $\chi$ is the chirality operator $\chi:= \ga_{2N+1} := 1_{\H_r}
\ox (-i)^N \ga^1 \ga^2 \dots \ga^{2N}$, and the real structure is the charge
conjugation for spinors $J := \left( 1_{\H_r} \ox (-i\chi)^N
\prod_{\nu=0}^{N-1} \ga^{2\nu+1} \right)\circ cc$, where $cc$ means complex
conjugation, for a representation of the $\ga$'s as in \cite{ZJ}.

In~\cite{nous} it is shown in full detail that Moyal planes are nonunital
spectral triples. In particular, the compactness condition is obtained
by complex interpolation between Hilbert-Schmidt and uniform norm
of $L^\th_f g(-i\nabla)$ for $f\in\A_\th$, $g\in L^p(\R^{2N})$ and a
suitable $p$. Indeed, the computation of the kernel of
$L^\th_f g(-i\nabla)$ and Lemma \ref{rep} yield
$\|L^\th_f g(-i\nabla)\|_2=(2\pi)^{-N}\|f\|_2 \|g\|_2$ and
$\|L^\th_f g(-i\nabla)\| \leq (2\pi\th)^{-N/2}\|f\|_2 \|g\|_\infty$.
Because the construction of the action functional is closely
related to the axiom of spectral dimension, only the latter will be tackled
here. Let us then begin by the main theorem.

\begin{thm}
\label{pr:calcul}
For $f \in \SS(\R^{2N})$, the compact operator
$\pi^\th(f)\,(|\Dslash| + \eps)^{-2N}$ lies in
$\L^{(1,\infty)}(\H)$ and any of its Dixmier trace $Tr_\omega$
is independent of~$\eps$. More precisely we have,
\begin{equation}
\Tr_\omega \bigl( \pi^\th(f)\,(|\Dslash| + \eps)^{-2N} \bigr)
= \frac{2^N\,\Omega_{2N}}{2N\,(2\pi)^{2N}} \int f(x) \,d^{2N}x
= \frac{1}{N!\,(2\pi)^N} \int f(x) \,d^{2N}x,
\label{eq:cojoformula}
\end{equation}
where $\Omega_{2N}$ is the hyper-area of the unit sphere in $\R^{2N}$.
\end{thm}
\noindent
 This result is exactly the same as in the commutative (compact or not) case
(see~\cite{ChakrabortyGS,ConnesAction,nous,Polaris}); this is the analogue
of Connes' trace Theorem~\cite{ConnesAction,Polaris} for the Moyal
flat space. In particular, the Dixmier trace is independant of the deformation
parameter $\th$. (However, as we see in the next section, the action
functional will depend on $\th$). Because $2N$ is the biggest exponent giving a
nonvanishing Dixmier trace, Theorem \ref{pr:calcul} yields that the spectral
dimension of the Moyal plane is $2N$. Moreover, this result confirm that
$Tr_\omega \bigl(\;.\;(|\Dslash| + \eps)^{-2N} \bigr)$ is a good abstractly defined
integral (see \cite{Book}).

Let first explain how the result~\eqref{eq:cojoformula} can be conjectured.
In view of the first axiom, we only need to work with a dense ideal of $\A_\th$,
and $\B_\th$ seems to be the most suitable (actually, we need integrable and
square integrable functions). By formula~(\ref{mogeneral}),
$\pi^\th(f)$ with $f\in\SS(\R^{2N})$ is a regularizing pseudodifferential
operator with symbol $\sigma[\pi^\th(f)](x,\xi)=f(x-{\textstyle
\frac{\th}{2}}S\xi)$. Now the symbol formula for a product of two $\Psi$DOs
yields
$$
\sigma\left[\pi^\th(f)\!(|\Dslash| + \eps)^{-2N}\right]\,(x,\xi)
=f(x-{\textstyle \frac{\th}{2}}S\xi)(|\xi| + \eps)^{-2N}\ox1_{2^N}.
$$
Let us define the regularized trace for a $\Psi$DO:
$$
\Tr_\La(A) := (2\pi)^{-2N} \iint_{|\xi|\leq\La}
\tr\left(\sigma[A](x,\xi)\right)
\,d^{2N}\xi \,d^{2N}x,
$$
where $\tr$ is the matricial part of the trace. The Dixmier trace is
heuristically linked (see~\cite{Langmann}) with $\lim_{\La\to\infty}
\Tr_\La(\.)/\log(\La^{2N})$. Doing this computation for
$A=\pi^\th(f)\,(|\Dslash| + \eps)^{-2N}$, we exactly find the right hand side
of~\eqref{eq:cojoformula}:
$$
\lim_{\La\to\infty}\frac{1}{2N\log\La}\Tr_\La \bigl(\pi^\th(f)
(|\Dslash| + \eps)^{-2N}\bigr)
= \frac{2^N\,\Omega_{2N}}{2N\,(2\pi)^{2N}} \int f(x) \,d^{2N}x.
$$
However, this is far from a rigorous proof; nor do we know how to mutate it
directly into one. The reader is invited to look up~\cite{nous} for a way
to establish it correctly, with regard to the Carey-Phillips-Sukochev
work~\cite{CareyPS}.

\section{The action functional}

Most of the interplays between the mathematical framework of noncommutative
differential geometry and physics come from computation of gauge actions.
These actions (functional and/or spectral) are intimately linked to the
spectral triples formalism. The Connes--Lott action functional, historically
the first, allows to deal with pure gauge systems (i.e., without gravity),
while the spectral action unifies gravitation with other interactions. In the
field of particle physics, one can recover with the action functional, in a
very natural way, the full Yang--Mills--Higgs sector of the standard model
(see~\cite{ConnesL1,CIS}). Here we compute the action functional associated
with the Moyal spectral triple, and reobtain the nowadays very used
noncommutative Yang--Mills action of the Moyal gauge
theory~\cite{Chaichian, DN, SMonster, Sz, Wul}.
So these theories naturally fit into the rigorous framework of spectral
triples. For simplicity, we will only concentrate on the $U(1)$ case, the
general Moyal-$U(n)$ being recovered with obvious modifications. We will first
introduce the last tools we need.

\subsection{The differential algebra}

Because of the close relation between the action functional and the spectral
dimension axiom, only $\B_\th$, the dense essential ideal of $\A_\th$
will be considered here. Let $\Omega^\8 \B_\th := \bigoplus_{p\in\N}
\Omega^p\B_\th$ be the universal differential graded algebra associated to
$\B_\th$. Recall that these graded algebra is generated by symbols:
$$
\Omega^n \B_\th := \Span\set{f_0\,\delta f_1 \dots \delta f_n},\;f_i\in\B_\th,
\sepword{for}\;i=1,\cdots,n \sepword{and}\;f_0\in\B_\th\oplus\C,
$$
and relations that can be deduced from properties of the universal
derivation:
$$
\delta(f\mop g)=\delta(f)  g+f \delta(g) \sepword{and} (\delta f)^* =\delta
f^*.
$$
With $\pi^\th$, one can represent $\Omega^\8 \B_\th$ on $\H$ by:
$$
\tpi^\th : \Omega^n\B_\th \to \L(\H) :
f_0\,\delta f_1 \dots \delta f_n \mapsto
i^n\,\pi^\th(f_0) \,[\Dslash,\pi^\th(f_1)]\dots[\Dslash,\pi^\th(f_n)],
$$
here we have implicitly extended $\pi^\th$ from $\B_\th$ to $\B_\th\oplus\C$.
Some unpleasant accidents may happened with such construction,
in particular represented forms can be the "image of zero", i.e.,
$\tpi^\th(\omega)=0$ while $\tpi^\th(\delta\omega)\neq 0$. To clear out
such forms,
Connes has introduced a two sided ideal of $\Omega^\8\B_\th$:
$\Junk := \bigoplus_{n\in \N} J^n$, $J^n:=J_0^n + \delta J_0^{n-1}$ where
$J_0^n := \Ker\left(\tpi^\th\left|_{\Omega^n\B_\th}\right.\right)$, and
finally we define,
$$
\Omega_\Dslash := \tpi^\th(\Omega^\8\B_\th)/\tpi^\th(\Junk).
$$

Thanks to the $\{f_{mn}\}$ basis, the 2-Junk (the only nontrivial
component needed here) is easily computable. In the next Lemma
we exhibit some particular elements of $\tpi^\th(J^2)$, with which
we are able to completely characterize $\tpi^\th(J^2)$.
\begin{lem}
\label{lm:basic-junk}
For $m,n,k,l \in \N^N$, let
$\omega_{mnkl} := f_{mk}\,\delta f_{kn} - f_{ml}\,\delta f_{ln} \in
\Omega^1\B_\th$ (no summation on~$k$ or~$l$). Then
$$
\tpi^\th(\omega_{mnkl}) = 0  \sepword{and}
\tpi^\th(\delta\omega_{mnkl})
= \tfrac{2}{\th}(|k| - |l|)\, L^\th_{f_{mn}} \ox 1_{2^N}.
$$
\end{lem}

\begin{proof}
Using the creation and annihilation functions~\eqref{eq:crea-annl},
and adopting the notations that, for $j = 1,\dots,N$, $\del_{a_j} =
\del/\del a_j$ and $\bar\del_{a_j} = \del/\del\bar a_j$, we
may rewrite the Dirac operator as follows;
$$
\Dslash = -\frac{i}{\sqrt{2}} \sum_{j=1}^N \ga^j(\del_{a_j} + \bar\del_{a_j})
+ i\ga^{j+N} (\del_{a_j}-\bar\del_{a_j})
= -i \sum_{j=1}^N (\ga^{a_j}\,\del_{a_j} + \ga^{\bar a_j}\,\bar\del_{a_j}),
$$
where $\ga^{a_j} := \frac{1}{\sqrt{2}}(\ga^j + i\ga^{j+N})$ and
$\ga^{\bar a_j} := \frac{1}{\sqrt{2}}(\ga^j - i\ga^{j+N})$.

Property~(\ref{poduit}), applied to $a_j$ and $\bar a_j$
respectively, yields
$$
\del_{a_j} = -\frac{1}{\th} \ad^{\mop}_{ \bar a_j}
:= -\frac{1}{\th} [\bar a_j,\.\,]_{\mop},  \qquad
\bar\del_{a_j} = \frac{1}{\th} \ad^{\mop}_{ a_j}
:= \frac{1}{\th} [a_j,\.\,]_{\mop}
$$
and hence,
$$
\Dslash = - \frac{i}{\th} \sum_{j=1}^N
(\ad^{\mop}_{a_j}\ox\ga^{\bar a_j} -  \ad^{\mop}_{\bar a_j}\ox\ga^{a_j}).
$$

Let $u_j := (0,0,\dots,1,\dots,0)$ be the $j$-th standard basis vector
of~$\R^N$. From definition \eqref{eq:basis}, we
directly compute:
\begin{align*}
\bar a_j \mop f_{mn} &= \sqrt{\th(m_j+1)}\, f_{m+u_j,n},
& f_{mn} \mop \bar a_j &= \sqrt{\th n_j}\, f_{m,n-u_j},
\\
a_j \mop f_{mn} &= \sqrt{\th m_j}\, f_{m-u_j,n},
& f_{mn} \mop a_j &= \sqrt{\th(n_j+1)}\, f_{m,n+u_j}.
\end{align*}
Consequently,
\begin{align*}
[\Dslash,\pi^\th(f_{mn})]
&= -\frac{i}{\th} \sum_{j=1}^N  \left(\sqrt{\th n_j}\, L^{\th}_{f_{m,n-u_j}}
- \sqrt{\th(m_j+1)}\,L^{\th}_{ f_{m+u_j,n}} \right) \ox \ga^{a_j}
\nonumber \\
&\hspace{4em} +  \left( \sqrt{\th m_j}\, L^\th_{f_{m-u_j,n}}
- \sqrt{\th(n_j+1)}\, L^\th_{f_{m,n+u_j}} \right)\ox\ga^{\bar a_j}.
\label{eq:plusun}
\end{align*}
Finally, using the projector or nilpotent properties of the $\{f_{mn}\}$
(Lemma \ref{fmn}), the result follows by a direct (quite long) computation.
\end{proof}

To identify $\tpi^\th(J^2)$, it suffices to remark that $\tpi^\th(J^2)
\subset \pi^\th(\B_\th)$. This follows because any
$\omega \in \tpi^\th(J^2) $ can be written as
$\omega = -\sum_{j\in I} L^\th_{\del_\mu f_j}
L^\th_{\del_\nu g_j} \ox \ga^\mu\ga^\nu$ where $I$ is a finite set, and
satisfies
$-i \sum_{j\in I} L^\th_{f_j \mop \del_\mu g_j} \ox \ga^\mu = 0$.
By the Leibniz rule,
\begin{align*}
\omega &= -\sum_{j\in I}\left(
L^\th_{\del_\mu(f_j \mop \del_\nu g_j)} -
L^\th_{f_j \mop \del_\mu\del_\nu g_j}\right)\ox \ga^\mu \ga^\nu
= \sum_{j\in I} L^\th_{f_j \mop \del_\mu \del_\nu g_j}
\ox \ga^\mu \ga^\nu\\
&= \sum_{j\in I} L^\th_{f_j \mop \del_\mu \del_\nu g_j}
\ox \eta^{\mu\nu}\,1_{2^N}.
\end{align*}
Because $\{f_{mn}\}$ is a basis for $\B_\th$, the previous Lemma yields
$\tpi^\th(J^2)\supset \pi^\th(\B_\th)$, so in the end we obtain:
\begin{prop}
\label{junk}
$\tpi^\th(J^2) = \pi^\th(\B_\th) = L^\th(\B_\th) \ox 1_{2^N}$.
\end{prop}

\subsection{The action}

In a canonical way, the functional or noncommutative Yang-Mills
action may be defined as:
\begin{equation}
\label{YM1}
YM(\a):= \frac{N!(2\pi)^N}{8g^2} \inf\set{I(\eta) : \tpi^\th(\eta) = \a}.
\end{equation}
The infimum is over all $\eta\in\Omega^1 \B_\th$ with the same image in
$\Omega^1_\Dslash$ of $I(\eta) := \Tr_\omega \bigl( \tpi^\th(F)^*
\,\tpi^\th(F)\, (\Dslash^2 + \eps^2)^{-N} \bigr)$, where
$\Omega^2 \B_\th\ni F=\delta\eta+\eta^2$ is the curvature of the universal
connection $\eta$. If we define $\Ht_n$ as the Hilbert space obtained by
completion of $\tpi^\th(\Omega^n \B_\th)$ under the scalar product
$$
\braket{\tpi^\th(\rho)}{\tpi^\th(\rho')}_n
:= \Tr_\omega \bigl( \tpi^\th(\rho)^* \,\tpi^\th(\rho')\,
(\Dslash^2 + \eps^2)^{-N} \bigr),
$$
for $\rho,\rho' \in \Omega^n \B_\th$, and $\H_n := P\Ht_n$ where $P$ be the
orthogonal projector on $\Ht_p$ whose range is the orthogonal complement of
$\tpi^\th(\delta J_0^{p-1})$, then one can show~\cite{ConnesL1,Sirius}:
\begin{equation}
YM(\a) = \frac{N!(2\pi)^N}{8g^2} \braket{P\tpi^\th(F)}{P\tpi^\th(F)}_2
\label{YM2}.
\end{equation}
With Theorem~\ref{pr:calcul} and Proposition~\ref{junk}, $YM(\a)$ in its
form~(\ref{YM2}) is easily computable, and one gets the expected result:

\begin{thm}
Let $\eta = -\eta^* \in \Omega^1\B_\th$. Then the Yang--Mills action
$YM(\a)$ of the universal connection $\delta + \eta$, with
$\a = \tpi^\th(\eta)$, is equal to
$$
YM(\a) = -\frac{1}{4g^2} \int F^{\mu\nu} \mop F_{\mu\nu}(x) \,d^{2N}x,
$$
where $F_{\mu\nu} :=
\half(\del_\mu A_\nu - \del_\nu A_\mu + [A_\mu,A_\nu]_{\mop})$ and
$A_\mu$ is defined by $\a = L^\th_{A_\mu} \ox \ga^\mu$.
\end{thm}
\noindent
We may remark that  property (\ref{trace}) yields:
$YM(\a)= -\frac{1}{4g^2} \int F^{\mu\nu}(x) \,F_{\mu\nu}(x) \,d^{2N}x$,
and because $F^*_{\mu\nu} = -F_{\mu\nu}$, $YM(\a)$ is
positive definite.

\section{Conclusions}

Apart from the mathematical questions raised in~\cite{nous} (like
the computation of the Hochschild cohomology of $\A_\th$ and $\Aun_\th$),
there are several remaining physical tasks. Firstly, it would
be desirable to extend our results to the generic Moyal
product~(\ref{mogeneral}).

On the other hand, a quite important job is to compute the spectral
action~\cite{AliAlain,CIKS} associated with this nonunital triple:
$S(\Dslash,A,a)=\Tr\left[\pi^\th(a)\phi_\Lambda\left(\Dslash_A^2\right)\right]$, where
$\phi_\Lambda$ is a suitable functional with a cutoff $\La$ and   $\Dslash_A=
\Dslash+A+JAJ^{-1}$, $A\in\tpi^\th\left(\Omega^1\B_\th\right)$. Here, it is the
definition of the spectral action in the nonunital case. The mathematical
problem is to compute a heat kernel (or other asymptotic~\cite{Odysseus})
expansion for a ``very nonminimal'' Laplace type operator, namely
the sqare of the ``Moyal-covariant'' Dirac operator $\Dslash_A$. The
treatment of quantization on Riemannian manifolds in~\cite{Klaas} should help.
Such computation will hopefully allow us to get a handle on noncommutative
gravity~\cite{ChamseddineGrav}.

\subsection*{Acknowledgments}

I would like to thank J.~C.~V\'arilly, J.~M.~Gracia-Bond\'{\i}a and my
Ph.~D.~advisor B.~Iochum for helpful discussions.

\end{document}